\documentclass[smallabstract,smallcaptions]{dccpaper}
\usepackage[utf8]{inputenc}
\usepackage{epsfig}
\usepackage{citesort}
\usepackage{amsmath}
\usepackage{amssymb}
\usepackage{color}
\usepackage{url}
\usepackage{tikz}
\usetikzlibrary{arrows}
\tikzstyle{block}=[draw opacity=0.7,line width=1.4cm]
\usepackage{qtree}
\usepackage{multicol}
\usepackage[lined, boxruled, linesnumbered]{algorithm2e}

\newlength{\figurewidth}
\newlength{\smallfigurewidth}

\begin{document}

\title
{\large
\textbf{Enumerative Data Compression with  Non-Uniquely Decodable Codes  }
}

\author{%
M. O\u{g}uzhan K\"{u}lekci$^{\ast}$, Yasin \"{O}zt\"urk$^{\ast}$,  Elif Altunok$^{\ast}$, Can Yılmaz Altıni\u{g}ne$^{\ast}$
\\[0.5em]
{\small\begin{minipage}{\linewidth}\begin{center}
\begin{tabular}{c}
Informatics Institute, Istanbul Technical University
\\
Bili\c{s}im Enstit\"us\"u, \.IT\"U Ayaza\u{g}a Kamp\"us\"u, Maslak,34469, \.Istanbul, Turkey
\\
\url{kulekci@itu.edu.tr} \\ 
\end{tabular}
\end{center}
\end{minipage}}
}

\maketitle
\thispagestyle{empty}

\begin{abstract}

Non-uniquely decodable codes can be defined as the codes that cannot be uniquely decoded without additional disambiguation information. These are mainly the class of non-prefix-free codes, where a codeword can be a prefix of other(s), and thus, the codeword boundary information is essential for correct decoding. Although the codeword bit stream consumes significantly less space when compared to prefix--free codes, the additional disambiguation information makes it difficult to catch the performance of prefix-free codes in total. Previous studies considered compression with non-prefix-free codes by integrating rank/select dictionaries or wavelet trees to mark the code-word boundaries. In this study we focus on another dimension with a block--wise enumeration scheme that improves the compression ratios of the previous studies significantly. Experiments conducted on a known corpus  showed that the proposed scheme successfully represents a source within its entropy, even performing better than the Huffman and arithmetic coding in some cases. The non-uniquely decodable codes also provides an intrinsic security feature due to lack of unique-decodability. We investigate this dimension as an opportunity to provide compressed data security without (or with less) encryption, and discuss various possible practical advantages supported by such codes.

\end{abstract}

\Section{Introduction}
A coding scheme basically replaces the symbols of an input sequence with their corresponding codewords. Such a scheme can be referred as non-uniquely decodable if it is not possible to uniquely decode the codewords back into the original data without using a disambiguation information. We consider non--prefix--free (NPF) codes in this study as the most simple representative of such codes. In NPF coding, a codeword can be a prefix of other(s), and  
the ambiguity arises since the codeword boundaries cannot be determined without explicit specification of the individual codeword lengths.

Due to the lack of that unique decodability feature, NPF codes has received very limited attention \cite{dalai2005,kulekci2013,adas2015} in the data compression area. Although the codewords become smaller when compared to their prefix-free versions, they should be augmented with the disambiguation information for proper decoding, and the additional space consumption of that auxiliary data structures unfortunately eliminates the advantage of short codewords.
Thus, the challenge here is to devise an efficient way of representing the codeword boundaries.

The data structures to bring unique decodability for NPF  codes was studied in \cite{kulekci2013}. 
More recently, the compression performance of  NPF codes, which are augmented with wavelet trees \cite{navarro2014wavelet} or rank/select dictionaries \cite{okanohara2007practical} to mark the code-word boundaries, had been compared with Huffman and arithmetic coding in \cite{adas2015}. It should be noted that using succinct bit arrays to mark the code-word boundaries had also been independently mentioned in some previous studies as well \cite{ferragina2007,fred2007}. Although such NPF coding schemes are performing a bit worse in terms of compression, they support random-access on compressed data.

In this work, we study improving the compression performance of non-uniquely decodable codes with the aim to close the gap with the prefix-free codes in terms of compression ratio. 
We propose an enumerative coding \cite{cleary1984comparison,kulekci2012enumeration} scheme to mark the codeword boundaries as an alternative of using wavelet trees or a rank/select dictionaries.
Instead of representing the length of every codeword on the encoded bit stream, the codeword boundaries are specified in blocks of $d$ consecutive symbols for a predetermined $d$ value. 
Assume the codeword lengths of the symbols in a block are shown with a $d$--dimensional vector. 
The sum of the $d$ individual codeword lengths is denoted by $p$, and the vector can be specified by its rank $q$ among all $d$--dimensional vectors having an inner sum of $p$ according to an enumeration scheme. 
Thus, a tuple $\langle p, q\rangle$, can specify the codeword boundaries in a $d$ symbol long block.

The method introduced in this study represents an input data by replacing every symbol with a NPF codeword and then compressing the corresponding $\langle p , q \rangle$ tuples efficiently.  
Experiments conducted on a known corpus \footnote{Manzini's corpus available at \url{http://people.unipmn.it/manzini/lightweight/corpus/index.html}.} showed that the compression ratios achieved with the proposed method reaches the entropy bounds and improve the arithmetic and Huffman coding ratios. To the best of our knowledge, this is the first study revealing that non-prefix-free codes can catch compression ratios quite close to entropy of the data. 


In recent years, compressive data processing, which can be defined as operating directly on compressed data for some purpose,  had been mentioned as a primary tool to keep pace with ever growing size in big data applications \cite{loh2012}. For instance, many database vendors are focusing on compressed databases  \cite{westmann2000implementation} to cope with the massive data management issues. 
On the other hand, it is becoming a daily practice to benefit from cloud services both for archival and processing of data. Obviously, the primary concern in using such a third-party remote service is the privacy and security of the data, which can be achieved simply by encryption. Encrypted compressed data is both space efficient and secure. However, the encryption level introduces several barriers in processing the underlying compressed data. Alternative solutions that investigate the privacy of the data without incorporating an encryption scheme have also been considered \cite{kulekci2012,kelley2014,gillman1996,rashmi2011}.  
Thus, new compression schemes respecting the data privacy without damaging the operational capabilities on the compressed data may find sound applications in practice. For instance,  similarity detection of documents without revealing their contents and privacy preserving storage with search capabilities are some potential applications based on those non-uniquely decodable codes.

The outline of the paper is as follows. We start by defining the non-prefix-free codes from a compression perspective, and then proceed by introducing our enumeration scheme to represent the disambiguation information. The proposed compression method as a whole is described next, which is then followed by the experimental results and discussions addressing the opportunities and future work. 

\Section{The Non-Prefix-Free Coding}
\label{sec::NPF}

\begin{figure}
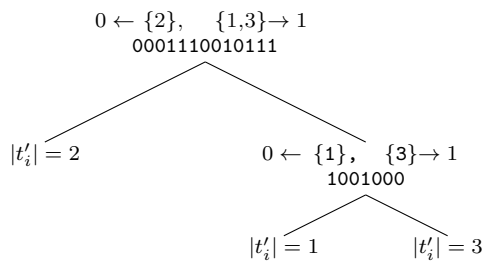

\scriptsize
\begin{center}

\begin{multicols}{2}

$T = NONPREFIXFREE$
\vspace{.2cm}

$\Sigma = \{E,R,F,N,I,O,P,X\}$

$O = \{ 3, 2, 2, 2, 1, 1, 1, 1\}$
\columnbreak

\renewcommand{\arraystretch}{1.2}
\begin{tabular}{rc|c|c|c|c|c|c|c}
\multicolumn{9}{c}{$\Sigma \rightarrow W $} \\
$\Sigma$: & E&R&F&N&I&O&P&X \\ \hline 
$W$ :& 0 & 1 &00 &01 &10 &11 &000 &001 \\
\end{tabular}
\end{multicols}

\texttt{$T \;=$ \fbox{{ }N}\fbox{{ }O}\fbox{{ }N}\fbox{{ }P{ }}\fbox{R}\fbox{E}\fbox{{ }F}\fbox{{ }I}\fbox{{ }X{ }}\fbox{{ }F}\fbox{R}\fbox{E}\fbox{E}} \\ 
\texttt{$T^\prime =$ \fbox{01}\fbox{11}\fbox{01}\fbox{000}\fbox{1}\fbox{0}\fbox{00}\fbox{10}\fbox{001}\fbox{00}\fbox{1}\fbox{0}\fbox{0}} \\
\texttt{$L \;=$ \fbox{{ }2}\fbox{{ }2}\fbox{{ }2}\fbox{{ }3{ }}\fbox{1}\fbox{1}\fbox{{ }2}\fbox{{ }2}\fbox{{ }3{ }}\fbox{{ }2}\fbox{1}\fbox{1}\fbox{1}} \\ 
\vspace{.2cm}
\textbf{a)} \textit{The $T^\prime = NPF(T)$ coding of a sample text $T$.}

\vspace{.4cm}

$\mathtt{T^\prime = 01110100010001000100100}$

$\mathtt{B \;\;= 10101010011101010010111}$
\vspace{.2cm}

\textbf{b)} \textit{Code-word boundaries in $T^\prime = NPF(T) $ marked on a bit array $B$.}

\vspace{.5cm}
{
\Tree [.{
\textsc{$0 \leftarrow$ \{2\}, \hspace{.2cm} \{1,3\}$\rightarrow 1$\ } \\
\texttt{0001110010111} 
} [.$|t^\prime_i|=2$ ] 
!\qsetw{5cm} [.{
\texttt{$0 \leftarrow$ \{1\}, \hspace{.05cm} \{3\}$\rightarrow 1$ } \\ 
\texttt{1001000}
} [.$|t^\prime_i|=1$ ] !\qsetw{3cm}
[.$|t^\prime_i|=3$ ] ]]
}
\vspace{.2cm}

\textbf{c)} \textit{Code-word lengths array $L$ in $T^\prime$ represented with a wavelet tree.}

\caption{The NPF coding and code-word boundary representation alternatives with bitmap and wavelet tree.}
\label{fig::NPF}
\end{center}
\end{figure}

$T=t_1t_2 \ldots t_n$ is a sequence of symbols, where $t_i \in \Sigma = \{ \epsilon_1, \epsilon_2, \ldots , \epsilon_\sigma \}$. Each symbol $\epsilon_i \in \Sigma$ requires  $\lceil \log \sigma \rceil$ bits in fixed-length coding, and the total length of $T$ then becomes  $n \cdot \lceil \log \sigma \rceil $ bits. Without loss of generality, assume  the symbols of the alphabet $\Sigma$ are ordered according to their number of occurrences on $T$  such that $\epsilon_1$ is the most and $\epsilon_\sigma$ is the least frequent ones.
 
Let's assume a  code word set $W=\{w_1, w_2, \ldots , w_\sigma\}$, where each $w_i$  denotes the minimal binary representation of $(i+1)$ as $w_i = MBR(i+1)$. 
The minimal binary representation of an integer $i >1$ is the  bit string 
$MBR(i) = b_1b_2\ldots b_{\log i}$  
 such that $i =  2^{\log i} + \sum_{a=1}^{\log i} b_a \cdot 2^{\log i - a} $. For example, $MBR(13)=101$, which is actually the binary representation of $13$ omitting the leftmost  $1$ bit.

This definition generates  $W = \{ 0, 1, 00, 01, 10, 11, 000, 001, \ldots \} $, where the code words  $w_i \in W$ has varying bit lengths, and $W$ is  not prefix free as some code words appear as the prefixes of others. The Kraft's inequality \cite{kraft46}, which  states that a code-word set $W$ is uniquely decodable if $\sum_{i=1}^{\sigma} 2^{-|w_i|} \leq 1$, does not hold on this $W$.
For each code-word length $\ell_k \in \{ 1, 2, 3, \ldots, \lfloor \log (\sigma + 1) \rfloor \}$, there are $2^{\ell_k}$ code words except the last code-word length by which less symbols might be represented when $\sigma \neq 2^h - 2$ for some $h$. Thus, it is clear that  $2^{-1} + 2^{-1} + 2^{-2} + 2^{-2}  + 2^{-2}  + 2^{-2}  + 2^{-3} \ldots + 2^{- (\lceil \log (\sigma + 2) \rceil -)1}  \geq 1$  when $\sigma >2$. 

The non-prefix-free coding of $T$ is the transformation obtained by replacing each  $ t_i = \epsilon_j $ with  $t^\prime_i = w_j$ according to the  $\Sigma \rightarrow W$ mapping  for all $1 \leq i \leq n$ as shown by  $NPF(T) = T^\prime =t^\prime_1t^\prime_2 \ldots t^\prime_n$, $t^\prime_i \in W$. 
In $T^\prime$, the most significant two symbols from $\Sigma$ are shown by 1 bit, and the following four symbols are denoted by 2 bits, and so on. 
The total number of bits in the non-prefix-free coded sequence $T$ is 
$|NPF(T)| = 1 \cdot(o_1 + o_2) + 2 \cdot (o_3+\ldots +o_6) + \ldots + (\lceil \log (\sigma + 2) \rceil -1) \cdot (o_{2^{\lceil \log (\sigma + 2) \rceil -1}-1  }+ \ldots + o_\sigma)$, where $o_i$ is the number of appearances of $\epsilon_i$ in $T$.         

The code word boundaries on $T^\prime$ are not self-delimiting and cannot be determined without additional information. 
Previous approaches \cite{kulekci2013,adas2015} used wavelet trees and rank/select dictionaries to mark the boundaries are shown in Figure \ref{fig::NPF}. 
Although these compressed data structures are very useful to support random access on the compressed sequence, it had been observed in  \cite{adas2015}  that the compression ratios achieved by these methods are a bit worse than the Huffman and arithmetic coding. 
In this study, we incorporate an enumerative coding to specify the codeword boundaries. 

Assume a list of items are ordered according to some definition, and it is possible to reconstruct any of the items from its rank in the list. In such a case, transmitting the index instead of the original data makes sense, and provides compression once representing the rank takes less space than the original data. That is actually the main idea behind enumerative coding \cite{cover1973enumerative}. We apply this scheme to represent the code-word boundaries in a sequence of NPF codewords. Empirical observations, as can be followed in the experimental results section, revealed that the usage of the proposed enumerative scheme can compress data down to its  entropy.

\Section{Enumerative Coding to Mark Codeword Boundaries}

One simple thing that can be  achieved to mark the codeword boundaries is to store the codeword lengths of individual symbols on the input text $T$. These lengths vary from minimum codeword length $1$ to a maximum of $\ell_{max} = \lfloor \log (\sigma +1) \rfloor$ bits. The sequence of $n$ codeword length information can then be compressed via a Huffman or arithmetic codec. However, our initial experiments showed that this coding does not provide a satisfactory compression ratio, where the total compression ratio cannot reach the entropy of the source sequence. With the motivation of marking the boundaries of multiple symbols instead of single individuals may improve the compression performance, we decided to test whether such a block-wise approach would help. 

A block is defined as consecutive $d$ symbols, and thus, there are $ r = \lceil \frac{n}{d} \rceil$ blocks on $T$. When $n$ is not divisible by $d$, we pad the sequence with the most frequent symbol. 
We maintain  a list of $r$ tuples as $R = \{\langle p_1, q_1 \rangle , \langle p_2, q_2 \rangle, \ldots \langle p_r, q_r \rangle\}$ such that 
\begin{itemize}
\item 
$p_i = |t^\prime_{(i-1)d+1}| + |t^\prime_{(i-1)d+2}|+ \ldots + |t^\prime_{i\cdot d}|$ for $1 \leq i \leq r$, where $|t^\prime_{j}| $ denotes the bit length of the codeword corresponding to symbol $t_j$, and 
\item 
$q_i$  represents the rank of the vector $\langle |t^\prime_{(i-1)d+1}|, |t^\prime_{(i-1)d+2}|, \ldots ,|t^\prime_{i\cdot d}| \rangle$ among all possible $d$-dimensional vectors whose elements sum up to $p_i$. 
\end{itemize}

For example on the example shown in Figure \ref{fig::NPF}, if we assume a block size of $d=3$, then $p_1 = 2 + 2 +2 = 6$ since the codeword lengths of the first three symbols (\texttt{NON}) are all $2$ bits. All possible $3$-dimensional vectors whose elements are in range $[1\ldots 3]$ and sum up to $6$ can be listed in lexicographic order as 
$\langle1,2,3\rangle$, 
$\langle1,3,2\rangle$, 
$\langle2,1,3\rangle$, 
$\langle2,2,2\rangle$, 
$\langle2,3,1\rangle$, 
$\langle3,1,2\rangle$, and 
$\langle3,2,1\rangle$. We observe that $\langle2,2,2\rangle$ is the fourth item in this list, and thus $q_1=4$. Similarly the lengths of the next block $\langle 3,1,1\rangle$ can be shown by $\langle 5, 6 \rangle$ since $3+1+1=5$ and $\langle 3,1,1\rangle$ is the sixth item in the lexicographically sorted possibilities list 
$\langle 1,1,3\rangle$,
$\langle 1,2,2\rangle$, 
$\langle 1,3,1\rangle$, 
$\langle 2,1,2\rangle$, 
$\langle 2,2,1\rangle$, and
$\langle 3,1,1\rangle$.

In such a block-wise approach we need to devise an enumeration strategy to convert an input vector to an index and vice versa. We explain the building blocks in the following subsections.

\SubSection{Number of Distinct Vectors}

Let $\psi(k,d,v)$ return the number of distinct $d$ dimensional vectors, in which each dimension can take values from $1$ to $k$, and they sum up to $v$ in total. The total sum $v$ should satisfy  $d \leq v \leq (k\cdot d)$ since each dimension is at least $1$ and at most $k$. If $v=d$ or $d=1$, then there can be only one possible vector in which all dimensions are set to 1 in the former case and to $k$ in the later case as there is only one dimension. The   $\psi(k,d,v)$ function can be computed with a recursion such that  $\psi(k,d,v) = \sum_{i=\alpha}^{\beta}  \psi(k,d-1,v-i)$. This is based on setting one, say first, dimension to one of the possible value $i$ and then counting the remaining $(d-1)$ dimensional vectors whose elements sum up to $(v-i)$. The pseudo code of this calculation is given in Algorithm \ref{algo:psi}.

\SubSection{Vector to Index}
Assume we are given a $d$ dimensional vector as $\langle v_1, v_2, \ldots ,v_d \rangle$, where each $1 \leq v_i \leq k$ for a known $k$. We would like to find the lexicographical rank of this vector among all possible $d$--dimensional vectors with an inner sum of $v=v_1+v_2+\ldots +v_d$. First we can count how many of the $d$-dimensional vectors have a smaller number than $v_1$ in their first dimension. Next step is to count the number of vectors that have the same $v_1$ in the first dimension, but less than $v_2$ in the second position. We repeat the same procedure on remaining dimensions, and the sum of the computed vectors return the rank of our vector. This can be achieved via a recursion, which is shown in Algorithm \ref{algo:vec2ind}, that uses the $\psi()$ function described above.
As an example, for $d=3$, and $k=3$, assume we want to find the rank of $2,2,2$. First we count the number of vectors that has a $1$ in its first position with an inner sum of $6$ via the $\psi(k=3,d=2,v=5)$ function, which returns us $2$. Next, we count  the vectors that has a 2 in first position and a value less than 2, which can take value only 1, in its second position. This can also be computed via  $\psi(k=3,d=1,v=3)$ function, which returns $1$ since we have only one dimension to set. Now we know that there are 3 vectors that are enumerated before our input on the possibilities list. We do not need to search for the last dimension since it is not free and its value is already determined.

\SubSection{Index to Vector}

In this case we are given a number $I$ representing the rank of a $d$ dimensional vector in a set of $d$ dimensional vectors with a known inner sum $v$, and we aim to generate this vector. We start by setting the first dimension to the minimum value $1$, and count how many possibilities exits by the $\psi(k,d-1,s-1)$. If this number is less than $I$, we decrease $I$ by this value, set $2$ for the first position and keep counting the possibilities in the same way until detecting the first value at which $I$ is no longer larger. Thus, we have found the first dimension of the vector, we repeat the same procedure to detect the other dimensions. The pseudo code of this calculation is given at  Algorithm \ref{algo:ind2vec}.

\begin{minipage}[t]{5cm}
\null 
\begin{algorithm}[H]
\scriptsize
\SetKw{KwDownTo}{down to}
\SetKw{KwAnd}{and}
\KwIn{\\
$k$: Maximum value of a dimension. \\
$d$: The number of dimensions.\\
$v$: The inner sum of the vectors. 
}
\KwOut{\\Number of distinct $d$ dimensional \\ vectors with an inner sum of $v$ .}
\lIf {$(v>k\cdot d) || (v<d)$} {\Return{0}}
\lIf {$(d=1) || (v=d)$} {\Return{1}}
\lIf {$(v=d+1)$} {\Return{d}}
\uIf{$(1<v+k-k\cdot d)$}{$\alpha = v+k-k\cdot d$}
\uElse{$\alpha = 1$}
\uIf{$(k<v-d+1)$}{$\beta = k$}
\uElse{$\beta = v-d+1$}
$sum=0$\;
\For{$(i=\alpha; i\leq \beta; i+=1)$}{$sum +=\psi(k,d-1,v-i)$\;}
\Return{$sum$};
\caption{$\psi(k,d,v)$ \label{algo:psi}}
\label{alg:psi}
\end{algorithm}
\end{minipage}
\begin{minipage}[t]{10cm}
\null 
\begin{algorithm}[H]
\scriptsize
\SetKw{KwDownTo}{down to}
\SetKw{KwAnd}{and}
\KwIn{
$k$: Maximum value of a dimension.
$d$: The number of dimensions.
$v_1 \ldots v_d$: Input vector. 
}
\KwOut{Rank of the input vector among lexicographically sorted vectors with  the same inner sum of $\sum_i v_i$ .}
$v = v_1 + v_2 + \ldots + v_d$ \;
\lIf {$(d=1) || (v=d)$} {\Return{0}}
$index=0$\;
\For{$(i=1; i< v_1; i+=1)$}{$index +=\psi(k,d-1,v-i)$\;}
$index +=$ VectorToIndex$(\langle v_2, v_3, \ldots, v_d \rangle, d-1,k)$\;
\Return{$index$};
\caption{VectorToIndex$(\langle v_1,v_2, \ldots,v_d \rangle,d,k)$ \label{algo:vec2ind}}
\label{alg:psi}
\end{algorithm}
\begin{algorithm}[H]
\scriptsize
\SetKw{KwDownTo}{down to}
\SetKw{KwAnd}{and}
\SetKwFunction{rankone}{\texttt{rank1}}
\SetKwFunction{rankzero}{\texttt{rank0}}
\KwIn{
$k$: Maximum value of a dimension
$d$: The number of dimensions.
$v$: The inner sum of the vectors.
$index$: The rank of the vector among all possible vectors.
}
\KwOut{The $\langle v_1,v_2, \ldots,v_d \rangle$ vector with $v_1+v_2+\ldots +v_d =v$, and rank $index$ among all possible vectors with inner sum $v$.}

\For{$(i=1; i<d; i+=1)$}
{
$v_i=1$\;
\While{$(z= \psi(k,d-i,v-v_i) < index)$}{
$index -= z$\;
$v_i=v_i+1$\;
}
$v = v-v_i$\;
}
$v_d = v$\;
\caption{$IndexToVector(k,d,v,index)$ \label{algo:ind2vec}}
\label{alg:psi}
\end{algorithm}
\end{minipage}

\Section{The Complete Method}

The pseudo-codes of the proposed encoding and decoding with the Non-uniquely decodable codes are given in Algorithms \ref{algo:encode} and \ref{algo:decode}. 

In the encoding phase, the NPF codeword stream $B$ is simply the concatenation of the NPF codewords. At each $d$ symbols long block, the total length of the codewords is the corresponding $p_i$ value, which can take values from $d$ to $k\cdot d$. This $p_i$ value is encoded to the $Pstream$ by an adaptive compressor. 

The index corresponding to the vector of the latest $d$ codeword lengths is computed with the $VectorToIndex$ function as described in the enumeration section by using the $p_i$ value as the inner sum. 
This $q_i$  is then encoded into the $Qstream$ with $p_i$ assumed to be the context in this compression. 
Notice that according to the $p_i$ value, the number of possible vectors change, where there appears relatively small candidates for small $p_i$. 
When all the codewords in the block are $1$ bit long, which means $p_i=d$, then there is no need to encode $q_i$
since there is only one possibility. Similarly, $p_i = k \cdot d$ implies all codeword are maximum length $k$, and again nothing is required to add into the compressed $Qstream$. 

The decoding phase is performed accordingly, where first the $p_i$ value is extracted from $Pstream$. 
If the extracted $p_i$ is equal to $d$ or $k \cdot d$, this implies nothing has been added to the $Qstream$ in the coding phase since the target vectors are determined with single options. Otherwise, by using the $p_i$ value as the context, 
the corresponding $q_i$ is extracted from $Qstream$ followed by the $IndexToVector$ operation. 

\begin{minipage}[t]{6.5cm}
\null 
\begin{algorithm}[H]
\scriptsize
\SetKw{KwDownTo}{down to}
\SetKw{KwAnd}{and}
\KwIn{
$T=t_1t_2\ldots t_n$ is the input data, where $t_i \in \Sigma=\{\epsilon_1, \epsilon_2, \ldots , \epsilon_\sigma\}$.  
$d$ is the chosen block length.\\
}
\KwOut{The codeword bit-stream and the compressed $\langle p_i,q_i \rangle$ list.}
$r = \lceil \frac{n}{d} \rceil$ \;
$B = \emptyset$ \;
Generate the NPF codeword set $W=\{w_1,w_2,\ldots,w_\sigma\}$\;
$k= \lfloor \log (\sigma+1) \rfloor$\;
\For{$(i=0; i<r; i+=1)$}{
$p_i= 0$\;
\For{$(j=0; j<d; j+=1)$}{
$\epsilon_h = T[i\cdot d + j + 1]$\;
$B \leftarrow Bw_h$\;
$vec[j+1] = |w_h|$\;
$p_i+=vec[j+1]$\;
}
Encode $p_i$ into $Pstream$ with an adaptive coder\;
\uIf{$(p_i \neq d) \&\& (p_i \neq k \cdot d)$}{
$q_i = VectorToIndex(vec[],d,k)$ \;
Encode $q_i$ into $Qstream$ with an adaptive coder by using the $sum$ value as the context\;
}
}
\caption{ \newline Encode$(T,d)$ \label{algo:encode}}
\label{alg:psi}
\end{algorithm}
\end{minipage}
\begin{minipage}[t]{7.5cm}
\null 
\begin{algorithm}[H]
\scriptsize
\SetKw{KwDownTo}{down to}
\SetKw{KwAnd}{and}
\KwIn{
$B$ is the NPF codeword bit stream. 
$Pstream$ is the compressed $p_i$ values. 
$Qstream$ is the compressed $q_i$ values.
$W=\{w_1,w_2,\ldots,w_\sigma\}$ is the NPF codeword set. 
}
\KwOut{The original data sequence $T=t_1t_2\ldots t_n$}
$r = \lceil \frac{n}{d} \rceil$ \;
$k= \lfloor \log (\sigma+1) \rfloor$\;
\For{$(i=0; i<r; i+=1)$}{
Decode $p_i$ from the $Pstream$\;
\uIf {$p_i=d$} {$\langle v_1, v_2, \ldots, v_d \rangle = \langle 1,1, \ldots,1\rangle$\;}
\uElseIf{$p_i=k\cdot d$}{$\langle v_1, v_2, \ldots, v_d \rangle = \langle k,k, \ldots, k\rangle$\;}
\Else{
Decode $q_i$ from the $Qstream$ by using $p_i$ as the context \;
$\langle v_1, v_2, \ldots v_d\rangle \leftarrow IndexToVector(k,d,p_i,q_i)$ \;
}
\For{$(j=1; j\leq d; j+=1)$}{
$w_h \leftarrow$ Read next $v_j$ bits from $B$\;
$t_{i\cdot d +j} = \epsilon_h$\;
}
}
\caption{\newline \small Decode$(B,Pstream,Qstream,d,n,W)$ \label{algo:decode}}
\label{alg:psi}
\end{algorithm}
\end{minipage}

\Section{Implementation and Experimental Results}

\begin{figure}
\begin{center}
\begin{tabular}{cc}
\includegraphics[scale=.4]{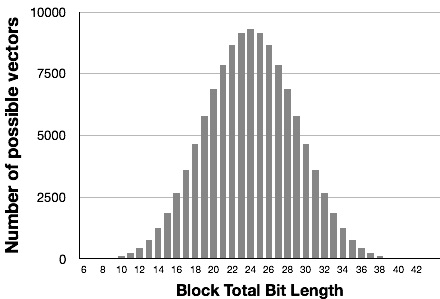} &
\includegraphics[scale=.4]{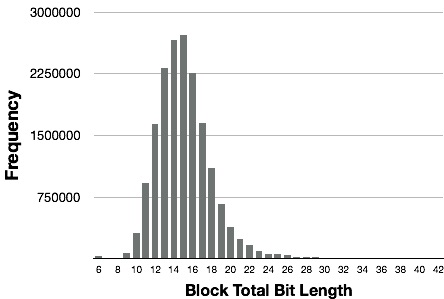} \\
a) & b) \\
\end{tabular}
\end{center}
\caption{\label{fig::pqgraphs} Assuming a block size of $d=6$ symbols, and a maximum codeword length $k=7$, a) presents possible bit block lengths and corresponding number of distinct vectors per each, b) presents the \emph{observed} bit block lengths and their number of occurrences on 100MB of English text (etext  file from the Manzini's corpus).}
\end{figure}

\begin{figure}
\begin{center}
\begin{tabular}{c}
\includegraphics[scale=.4]{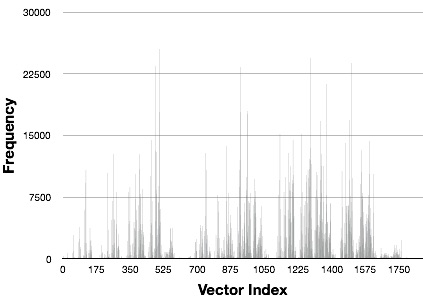} 
\end{tabular}
\end{center}
\caption{\label{fig::qi} The distribution are $1875$ distinct $6$--dimensional vectors, where each dimension can take values from $1$ to $k=7$ with an inner sum of $15$ on 100MB of English text according to our enumeration scheme.}
\end{figure}

Being directly proportional to the imbalance of the symbol frequencies in the source, the codewords with short lengths are expected to appear more, and thus, the $d$--dimensional vectors are in general filled with small numbers with small $p_i$ values as a consequence. Figure \ref{fig::pqgraphs} shows the distribution of block lengths and their corresponding number of distinct vectors by assuming $d=6$ and $k=7$. On the same figure also the observed frequencies of possible block bit lengths on a 100 megabyte English text are depicted, where the most frequent bit block length seems $15$ here. We present the distribution of $q_i$ values in the context of $p=15$ on Figure \ref{fig::qi} to give an idea about the imbalance that increases the success of representing codeword boundaries over the non--prefix--free codeword stream. 

We have implemented the proposed scheme and compared  compression ratio against both static and adaptive versions of the Huffman and arithmetic codes (AC) on the test corpus.  
While compressing the $Pstream$ and $Qstream$, we have used the adaptive arithmetic encoder of \cite{said2004}, and tested our scheme with different block sizes of $d=2$, $d=4$, and $d=6$.  
Table \ref{tab::res} shows the compression performance of each scheme on various files in terms of bits spent per each symbol in total.

The experiments showed that for $d=6$, the compression ratio of the Non-uniquely decodable codes generally improves the others.  
However, AC seems achieving better ratios on three files. On \texttt{howto} file the difference in between the AC and Non-uniquely decodable codes are very small as being in thousands decimal, which is not found to be meaningful. 
On \texttt{howto.bwt}, which is the same \texttt{howto} file after Burrows-Wheeler transform, the difference is sharper. This is mainly due to the fact that the runs in the BWT string may introduce an advantage for the adaptive codes. Notice that both files are around $40$ megabytes and shorter than the other files except the \texttt{chr22.dna} file, on which our method performs clearly worse. In the current experimental observations it is thought that the performance of the proposed coding becomes better on large files with larger alphabets. 

It is possible to increase the block size, particularly on larger volumes. However, when $d$ becomes larger current implementation suffers from the slow down due to the recursive function implementations to find the enumerative index of a vector, and vice versa. Considering that our compression scheme is composed of three main components as the base non-prefix-free code stream, and over that the $Pstream$ and $Qstream$, we would like to monitor their respective space occupation on the final compressed size. Table \ref{tab::subanalysis} includes the diffraction of these three components for different $d$ values tested. There appears a trade off such that the $Pstream$ gets better compressed with increased block size, where the reverse works for $Qstream$. 

\begin{table}
\scriptsize
\begin{tabular}{lrr|c|cc|cc|cc|ccc}
        &          &            &           & \multicolumn{2}{c}{Huffman} & \multicolumn{2}{|c|}{Arithmetic} & \multicolumn{2}{c|}{NPF} & \multicolumn{3}{c}{Non-uniquely decodable} \\ 
File	&	Size	&	Symbols	&	Entropy	&	Stat.	&	Adapt.	&	Stat.	&	Adapt.  & RS    & WT 	&	d=2	&	d=4	&	d=6	\\ \hline
sprot34.dat	&	109MB	&	66 (k=6)	&	4.762	&	4.797	&	4.785	&	4.764	&	4.749   & 5.434 & 5.178 	&	4.869	&	4.790	&	\textbf{4.698}	\\ \hline
chr22.dna	&	34MB	&	5 (k=2)	    &	2.137	&	2.263	&	2.195	&	2.137	&	\textbf{1.960}  & 2.957     &   2,616 	&	2.468	&	2.466	&	2.462	\\ \hline
etext99	&	105MB	&	146 (k=7)	    &	4.596	&	4.645	&	4.595	&	4.604	&	4.558	& 5.140 & 4,553     &	4.632	&	4.570	&	\textbf{4.553}	\\ \hline
howto	&	39MB	&	197 (k=7)	    &	4.834	&	4.891	&	4.779	&	4.845	&	\textbf{4.731}	& 5.300 	&   4.215   &   4.856	&	4.759	&	4.736	\\ \hline
howto.bwt	&	39MB	&	198 (k=7)	&	4.834	&	4.891	&	3.650	&	4.845	&	\textbf{3.471}	& 5.300     &   4.215   &   4.143	&	3.950	&	3.949	\\ \hline
jdk13c	&	69MB	&	113 (k=6)	    &	5.531	&	5.563	&	5.486	&	5.535	&	5.450	&	6.404   & 5.658 &   5.577	&	5.460	&	\textbf{5.275}	\\ \hline
rctail96	&	114MB	&	93 (k=6)	&	5.154	&	5.187	&	5.172	&	5.156	&	5.139	&	5.766   & 5.408	&	5.164   &   5.020	&	\textbf{4.818}	\\ \hline
rfc	&	116MB	&	120	(k=6) &	4.623	&	4.656	&	4.573	&	4.626	&	4.529	&	5.094   &   4.853   & 4.685	&	4.555	&	\textbf{4.463}	\\ \hline
w3c2	&	104MB	&	256 (k=8)	    &	5.954	&	5.984	&	5.700	&	5.960	&	5.659	&	6.648   & 5.820 &   5.826	&	5.686	&	\textbf{5.617}	\\ 
\end{tabular}
\caption{\label{tab::res} Compression ratio comparison between the proposed scheme, NPF rank/select and wavelet tree \cite{adas2015}, arithmetic, and Huffman coding in terms of bits/symbol.}
\end{table}

\begin{table}
\begin{center}
\scriptsize
\begin{tabular}{c|c|ccc|ccc}

	    &	Codeword	&	\multicolumn{3}{c}{Pstream} 	&	\multicolumn{3}{|c}{Qstream} 			\\	
File	&	Stream	    &	d=2	&	d=4	&	d=6 	&	d=2	&	d=4	&	d=6 	\\	\hline
sprot34.dat	&	2.686	&	1.476	&	0.909	&	0.659	&	0.707	&	1.196	&	1.353	\\	\hline
chr22.dna	&	1.494	&	0.718	&	0.504	&	0.399	&	0.256	&	0.468	&	0.568	\\	\hline
etext99	&	2.516	&	1.316	&	0.789	&	0.580	&	0.800	&	1.265	&	1.457	\\	\hline
howto	&	2.618	&	1.451	&	0.885	&	0.655	&	0.787	&	1.256	&	1.464	\\	\hline
howto.bwt	&	2.618	&	1.183	&	0.781	&	0.604	&	0.342	&	0.552	&	0.726	\\	\hline
jdk13c	&	3.263	&	1.449	&	0.871	&	0.642	&	0.866	&	1.327	&	1.370	\\	\hline
rctail96	&	2.878	&	1.462	&	0.893	&	0.659	&	0.824	&	1.250	&	1.281	\\	\hline
rfc	&	2.516	&	1.472	&	0.911	&	0.677	&	0.697	&	1.128	&	1.271	\\	\hline
w3c2	&	3.436	&	1.548	&	0.949	&	0.706	&	0.841	&	1.301	&	1.475	\\	\hline
\end{tabular}
\caption{\label{tab::subanalysis} The diffraction of the codeword stream, Pstream, and Qstream on the number of bits used per symbol for different $d$ values. }
\end{center}
\end{table}

\Section{Discussions and Conclusions}

This study has shown that non-prefix-free codes with an efficient representation of the codeword boundaries can reach the entropy bounds in compression as is the case for prefix--free codes. 
The incentive to choose Non-uniquely decodable codes as an alternative to Huffman or arithmetic coding might be their intrinsic security features. It is not possible to decode the codeword stream without the codeword boundary information encoded in $Pstream$ and $Qstream$. Thus, there is no need to encrypt the codeword stream when one would like to secure the compressed data. More than that, it is still possible to make some operations such as search  and similarity computations on the codeword stream. Yet another opportunity might appear in the distributed storage of the data, where keeping the NPF codewords and codeword boundary informations in different sites can help in providing the security. Same idea may also apply in content delivery networks.

Besides the compression ratio, where this study mainly concentrated, the memory usage and the speed of compression are surely important parameters in practice. Current implementation is slow due to two main facts as NPF codewords are not byte--aligned, and the vector to/from index enumerations are consuming additional time. The former problem is common to all variable length codes, which can be overcome by benefiting from the Huffman coding tables idea \cite{sieminski1988fast}. The enumeration time consumption can also be decreased by using tables which include the precomputed vector to/from index calculations by sacrificing a bit more memory. The algorithm engineering of the proposed scheme along with the possible applications in data security area are possible venues of research as a next step. Surely, better data structures to encode the codeword boundaries is open for improvement. 

\Section{References}
\bibliographystyle{IEEEtran}
\bibliography{refs}

\end{document}